\documentstyle[prb,aps,multicol]{revtex}
  \renewcommand{\narrowtext}{\begin{multicols}{2} \global\columnwidth20.5pc}
  \renewcommand{\widetext}{\end{multicols} \global\columnwidth42.5pc}
  \newcommand{\wide}{\widetext \noindent \line(200,0){245} \line(0,1){3}\\}
  \newcommand{\narrow}{\begin{flushright}\mbox{\line(0,-1){3}$\! \!$
        \line(1,0){245}} \end{flushright} \narrowtext \noindent}

\newcommand{\br}{{\bf r}}
\newcommand{\bn}{{\bf n}}
\draft

\begin{document}
\title{Nonlinear acoustic and microwave absorption in disordered semiconductors}
\author{M. Kirkengen$^{a}$ and Y. M.Galperin$^{a,b,c}$}
\address{$^a$Department of Physics, University of Oslo,  P. O. Box 1048  
Blindern, N 0316 Oslo, Norway;\\
$^b$Centre for Advanced Studies, Drammensveien 78, 0271 Oslo,
Norway;\\
$^c$Solid State Division, A. F. Ioffe  
Physico-Technical  Institute, 194021 St. Petersburg, Russia}
\date{\today}

\maketitle
\begin{abstract} 
Nonlinear hopping absorption of ultrasound and electromagnetic waves
in amorphous and doped semiconductors is considered.  It is shown that
even at low amplitudes of the electric (or acoustic) field the
nonlinear corrections to the relaxational absorption appear
anomalously large. The physical reason for such behavior is that the
nonlinear contribution is dominated by a small group of close impurity
pairs having one electron per pair. Since the group is small, it is
strongly influenced by the field.  An external magnetic field strongly
influences the absorption by changing the overlap between the pair
components' wave functions. It is important that the influence is
substantially different for the linear and nonlinear
contributions. This property provides an additional tool to extract
nonlinear effects.
\end{abstract}


\narrowtext

\section{Introduction}

The subject of this article is nonlinear microwave and acoustic
properties of amorphous semiconductors and lightly doped crystalline
semiconductors in the regime of hopping conductance. We are interested
in absorption %of oscillations due 
to electron transitions between
localized states associated with defects or impurity atoms. We
consider the case where the absorption is due to electron hopping
within pairs of {\em neighboring} defects containing one electron per
pair.  The distance between the centers within the pair
must be small enough to allow tunneling, while the distance to other
impurities should be large enough to prevent tunneling to impurities
outside the pair. In a weakly doped semiconductor we can expect these
pairs to be relatively rare, and triplets of the same kind will be
even less likely. Thus a natural basis to treat the problem is the
so-called two level approximation according to which only the lowest
energy level of each of two neighboring impurities are taken in
consideration. This approach, as well as its range of applicability,
was first discussed in detail by Pollack and Geballe.\cite{PollGeb}

For brevity, in the following we shall discuss the case of acoustic
attenuation and then specify what changes in the formulae should be
introduced to allow for electromagnetic absorption.  

An external AC electric or acoustic field causes transitions between
the electron states.
Direct inter-level
transitions leading to absorption of quanta give rise to the so-called
\emph{resonant} absorption.  For low intensities, the resonant
contribution to the absorption coefficient of an acoustic wave can be
expressed as\cite{GGPRev}
\begin{equation} \label{GammaRes}
\Gamma^{(\text{res})} =\alpha_1 \omega/s\, \tanh \left(
\hbar \omega / 2 k_B T\right)
\end{equation}
where $\alpha_1$ is a dimensionless coupling parameter, weakly
dependent on temperature and frequency, $s$ is the sound velocity,
$\omega$ is the sound frequency, and $T$ is the temperature.  We can
always assume the relation $\hbar \omega / 2 k_B T \ll 1$.  $\alpha_1$
will be specified later, see Eq. (\ref{alpha}).  Defining $n_0$ as the
Fermi function $n_0(E)=[\exp(E/k_BT)+1]^{-1}$, we can write the factor
$\tanh \left(\hbar\omega/2 k_B T\right)$ as $n_0(\hbar \omega/2) -
n_0(-\hbar\omega/2)$. This can be recognized as the difference in the
equilibrium population of the two levels of the pair with an energy
splitting of $\hbar \omega/2-(-\hbar \omega/2)=\hbar \omega$, that is,
the pairs which can directly absorb a phonon.

The \emph{relaxation} absorption is due to a modulation of the
electron inter-level spacing $2\epsilon$ by the AC field. Such a
modulation leads to a periodic change of the occupation
numbers of the two levels which lags in phase the variation of
$\epsilon$. This lag leads to the energy dissipation.
In the linear regime the coefficient of relaxation absorption has been
calculated as\cite{Jackle}
\begin{equation} \label{GammaRel}
\Gamma^{\text{(rel)}} \approx \frac{\alpha_2}{s}
\left\{ \begin{array}{ll}\tau^{-1},& \omega \tau_0 \gg 1\\ \omega, &
\omega \tau_0 \ll 1 \end{array} \right. 
\end{equation}
where $\alpha_2 \approx \alpha_1$ and $\tau_0$ represents a minimal
relaxation time for $\epsilon \approx k_BT$. The physical meaning
of $\tau_0$, as well as estimates of this quantity,  will be
%presented more closely in our discussion of the relaxation time in
%the theoretical analysis. 
will be discussed later.

Comparing Eq.~(\ref{GammaRel}) with Eq.~(\ref{GammaRes}) we conclude that
 the relaxation absorption always predominates at $\omega \tau_0 \ll 1$. If
  $\omega \tau_0 \gg 1$ the ratio
$\Gamma^{\text{(res)}}/\Gamma^{\text{(rel)}}\approx \omega \tau_0
\tanh(\hbar \omega/2k_BT)$ can be either greater or less than one
under experimentally accessible conditions.

For higher intensities this comparison is no longer valid, as both the
resonant and the relaxation absorption show strongly nonlinear
behavior. For the resonant absorption, the nonlinearity is due to an
equalization of the population numbers of the two electron
states. {}From Eq.~(\ref{GammaRes}) we see that this leads to a strong
reduction of the resonant absorption. Usually this suppression takes
place at very low intensities.\cite{Jackle}

The nonlinearity of the relaxation absorption is due to the
following. In the limit of high intensities, when the perturbing
potential is amplitude $\gtrsim k_BT$, there will be times when
 $2 \epsilon(t) \le k_B T$ and  both 
states are almost equally occupied. During such part of the wave period no
transitions and thereby no absorption will occur. Consequently, the
absorption coefficient decreases with the sound amplitude. 

The linear relaxation absorption in semiconductors has previously been
studied both theoretically\cite{ESAbs} and experimentally (for a
review, see Ref.~\onlinecite{Long}), while for the strongly nonlinear
regime only theoretical results\cite{GGPRev} were obtained.  The
reason why this type of nonlinearity has not been observed
experimentally is probably some masking by other mechanisms leading to
a nonlinear behavior. One of such mechanisms could be wave-induced
ionization of impurity atoms into the conduction band observed in
InSb.\cite{GGDL} Consequently, to observe the mechanism\cite{GGPRev}
of nonlinear absorption one should carefully choose the material. That
does not seem to be an easy task, and we are not aware of any
experiments of this type.

In this article we will address the case of \emph{low} intensities when
nonlinear effect manifest themselves as small corrections to the
linear absorption. The main message is that
already the lowest order corrections offer interesting information and
anomalous effects worth studying. Hopefully, these effects will be
pronounced within an experimentally achievable parameter space where
other mechanisms of nonlinear behavior are still not important. 

The two-site approximation for semiconductors allows us to describe 
our system in terms of the so-called Two Level System (TLS) model.
This model was first proposed independently 
by Anderson {\em et al.}\cite{AHV} and Philips \cite{Phil}
to explain the low temperature specific heat in amorphous 
materials, and has been successful in describing several other
phenomena. The model has given a theoretical explanation for a 
surprising universality in the behavior of very different disordered
materials at low temperatures. 
 
The authors\cite{KG} have previously used the TLS model to analyze nonlinear
corrections to the absorption in both dielectric and metallic
glasses. However, there are several important differences between the
situations in glasses and in disordered or doped semiconductors. 
The main difference is due to
the electric charge of the particle involved in the transitions. This
introduces two modifications to the earlier results.

First, the electric charge gives the pair a dipole moment proportional
to the distance between the pairs, the coupling being proportional to
the dipole moment's component along the direction of the electric
field. This leads to a specific orientational dependence of the
absorption of one pair, and after integration over all pairs it
significantly influences the absorption.  Second, we must expect our
system to behave differently with the application of a magnetic
field. More specifically, as the magnetic field leads to a stronger
localization of the electrons, we must expect the absorption to
decrease with an applied field. The effect of a magnetic field will
therefore also be analyzed, in the special cases of a weak or a strong
field. The connection between the orientation of the dipole moment and
the magnetic field direction also leads to a dependence of the
absorption coefficient on the relative directions of the radiation
wave vector and the magnetic field. We will solve the problem for a
magnetic field parallel or perpendicular to the wave vector of the
radiation. 
%, as all other directions can be seen as linear combinations of the two.

The magnetic field dependence gives us the possibility to separate the
relaxational absorption from other contributions. It will also be
shown that the effect of a magnetic field depends on the specific
relaxation mechanism, thus providing us with a tool to further
understand the relaxation %absorption.
processes for localized carriers.

The paper is organized as follows.  First, we give a short
introduction to the theory based on the TLS model and how it can be used to
solve the problem of relaxation absorption in glasses. Then we will fit
the two-site approximation for a semiconductor to the TLS model and
show what modifications are needed for this. Finally, we will show how
the effects of a magnetic field can readily be included in the
analysis, and what effects to expect from this.

\section{%Theoretical Background, Two Level Systems}
Two Level Systems and relaxation losses in glasses}
The TLS model deals with a particle (in our case, an electron) moving
in a slightly asymmetric double well potential. It is assumed that
only the two ground levels are accessible. 
The ground levels in the
isolated wells are assumed to have a slight separation in energy,
$2\Delta$, and to be coupled via a tunneling energy overlap integral
$\Lambda$. The Hamiltonian of such a system 
is traditionally written as
\begin{equation} \label{Hamil0}
{\cal H}_0=\Delta \, \sigma_z - \Lambda\, \sigma_x
%\left[ \begin{array}{cc}\Delta & -\Lambda \\-\Lambda
%&-\Delta\end{array} \right]
\end{equation}
where $\sigma_i$ are the Pauli matrices.

Let us now apply an external periodic perturbing potential and study
the power absorbed by a single pair, $p(\Delta,\Lambda)$. As is will
be shown, this power is in general a non-monotonous function of its
parameters, and there exists an ``optimal'' region which dominates
absorption. What is important is that those regions are
\emph{different} for the linear absorption and for the nonlinear
correction. Let us define $\delta$ as the typical value of the
inter-level splitting $\Delta$ which is important for the onset of
nonlinear behavior. Estimates for the quantity
$\delta$ will be given in the discussion section.

At $\hbar \omega \ll \delta$ one can employ the adiabatic
approximation and neglect time derivatives of the external field while
solving the Schr\"odinger equation for the TLS.\cite{Galdiel} In this
approximation we write the interaction Hamiltonian as\cite{Jackle}
\begin{equation} \label{HamilI}
{\cal H_I}=\sigma_z\, d \cos \omega t\, 
%\left[ \begin{array}{cc}1 & 0\\0 & -1\\\end{array}\right]
\end{equation}
ignoring possible off-diagonal items.\cite{Jackle,Black} The quantity
$d$ is just the coupling constant between the field and the TLS. In
the physics of low-temperature properties of glasses $d$ is assumed to
be a random quantity, uncorrelated with $\Delta$ and $\Lambda$. This
assumption is generally not valid for the case of semiconductors.

In the case of a sound wave $d=\gamma_{ik}u_{ik}^{(0)}$, where
$\gamma_{ik}$ is the deformational potential of the TLS and
$u_{ik}^{(0)}$ is the amplitude value of the deformation tensor.

For semiconductors the value of $\Lambda$ depends on the spatial
separation of the wells. In the case of electromagnetic waves $d=\eta
{\cal E}_0$, where $\eta$ is the dipole moment of the TLS while ${\cal
E}_0$ is the amplitude of the electric field.\cite{Jackle,HA} For
charged particles the dipole moment is proportional to the distance
between the wells, and is thereby strongly correlated with the value
of $\Lambda$.
 
The total Hamiltonian of the TLS may now be written as\cite{Jackle}   
\begin{equation}\label{Hamil}
{\cal H}=(\Delta + d\, \cos \omega t)\, \sigma_z -\Lambda \, \sigma_x\,.
\end{equation}
The difference  between the eigenvalues of this new Hamiltonian,
$2\epsilon (t)$, where
\begin{equation} \label{esplitting}
\epsilon (t) = \sqrt{(\Delta+d \cos \omega t)^2 + \Lambda ^ 2}\, .
\end{equation}
 is just the energy splitting of the TLS.\cite{end1}
To characterize a TLS we need not only the energy spa\-cing, but also 
the occupation numbers of the upper ($n$) and lower ($1-n$) levels.
The non-equilibrium occupation numbers can be found from the balance
equation\cite{Galdiel}
\begin{equation} \label{nbalance}
\frac{dn}{dt} = -\frac{n-n_0(t)}{\tau (t)} 
\end{equation} 
where $n_0(t)$ is the adiabatic equilibrium occupation
number. It depends on 
the energy spacing $\epsilon (t)$, temperature $T$ and time $t$ as  
\begin{equation} 
n_0 (t) = \left[e^{2\epsilon (t)/k_BT}+1 \right]^{-1}.
\label{n_0}
\end{equation}
The relaxation time $\tau (t)$ is a function of the energy splitting, 
the tunneling barrier and temperature, and also depends on the exact 
relaxation mechanism. 

The power absorbed by a single TLS can be determined by the
expression\cite{Galdiel} 
\begin{equation} \label{absoneTLS}
p(\Delta, \Lambda)  =  \frac{2}{\Theta}\int_0^{\Theta}\!dt\,
n(t) \, \frac{ d
\epsilon}{d t} \, , \quad \Theta \equiv \frac{2\pi}{\omega}\, . 
\end{equation}

The contributions of individual TLS must be added and such a summation
can be performed in a conventional way using the distribution function
$N(\Delta, \Lambda)$ of the random parameters $\Delta$ and $\Lambda$
and replacing the deformational potential $\gamma_{ik}$ by its average value. 

To analyze the nonlinear absorption we use the exact periodic in
time solution of Eq.~(\ref{nbalance}) to obtain the following result
for the total absorbed power,\cite{Galdiel}
\begin{eqnarray} \label{esol}
P &=& \frac{1}{\Theta}
\int_0^\infty \!
\int_0^\infty \!  \frac{ N(\Delta,\Lambda) \, d
\Delta \, d \Lambda}{ k_B T} 
%\frac{1}{
\left(1 - e^{ -\int_0^{\Theta}
dt_1/\tau (t_1)}\right)^{-1}
%}
\nonumber \\ &\times &
\int_0^{\Theta} \!
\int_0^{\Theta} \! \frac{dt\, dt'\,{\dot \epsilon} (t){\dot
\epsilon}(t-t')}{\cosh^2 [\epsilon (t-t')/k_B T] } 
e^{- \int_0^{t'} dt_1/\tau
(t-t_1)} .
\end{eqnarray}

Unfortunately, this integral cannot be calculated analytically in the
general case.  Earlier discussions of the strongly nonlinear regime
have considered situations where only very limited time or energy
ranges contribute to the absorption. This corresponds to the
asymptotic high intensity limits of the integral. Our approach has
rather been expanding the integral in powers of the amplitude of the
modulation of the energy splitting, $d$, to get insights into the
first onset of nonlinear effects. So we will concentrate on the regime
of weak nonlinearity.

\section{Semiconductors in the TLS formalism}

We will now show how hopping in amorphous or doped crystalline
semiconductors can be described by the TLS model. 

A TLS is naturally formed by a pair of nearest impurity centers having
one electron per pair. To make the calculations for an individual pair
we have to specify the coupling constant $d$ and the relaxation time
$\tau$. The latter is essentially dependent on the dominant mechanism
of pair population relaxation. To sum over all the pairs we have to
specify the proper distribution function for the parameters of the
pairs containing one electron.

We will concentrate on the case when the interaction has a dipole
character and can be expressed in the form
\begin{equation} \label{ip2}
d (\br,t) = d_0(\br) \cos \omega t\, , \quad d_0 (\br) \equiv e{\cal
E}_0 r (\bbox{\nu} \cdot \bn)
\end{equation}
where ${\cal E}_0$ is the amplitude of the effective electric field
acting upon the electrons. It is just
the amplitude of the local 
electric field created either by the external electric field, our due
to piezoelectric interaction with an acoustic wave.\cite{end2}
$\bbox{\nu}=\bbox{\cal E}/{\cal E}$ is the field
polarization vector,  $r$ is the distance between the components
of the pair, $\bn =\br/r$ is the pair direction vector.     

The relaxation time is strongly dependent on the particular
mechanism  of interaction between localized electrons and phonons, see
for example review Ref.~\onlinecite{GGPRev}. Here we will only quote the
most important results. 
\subsection{Relaxation time}
In general, the relaxation time can be expressed as
\cite{GGPRev}
\begin{eqnarray} \label{rt}
\frac{1}{\tau(\epsilon,{\bf r})}& =& 
\frac{1}{\tau_n(T)} 
\left(\frac{\epsilon}{k_BT}\right)^n 
\left(\frac{\Lambda({\bf r})}{\epsilon }\right)^2 \nonumber \\ && \times
\Phi_n \left(\frac{2\epsilon }{E_r}\right) 
\frac{\coth(\epsilon/k_BT)}{[1+(2\epsilon/E_a)^2]^4},
\end{eqnarray}
where the exponent $n$, the factor $\tau_n^{-1}$ and the function
$\Phi_n(x)$ are dependent on the particular mechanism of interaction
between localized electrons and phonons. The meaning of the energies
$E_r$ and $E_a$ will be made clear in a moment. 

The energy dependence of $\tau^{-1}$ can be easily understood. The
phonon emitted or absorbed during a transition between the electron
levels must obviously have an energy equal to the energy splitting of
the pair, $2\epsilon$. The power by which it occurs in the formula
(\ref{rt}) is determined by the product of the phonon density of
states with the frequency dependence of the squared interaction matrix
element.  The factor $\coth(\epsilon/k_BT)$ is equal to
$2N_{\omega}+1$ where $N_{\omega}$ is the Planck function for $\hbar
\omega= 2\epsilon$. Apart from a proportionality factor, this is the
probability of phonon emission, $\propto N_{\omega}$, plus the
probability of absorption, $\propto (N_{\omega}+1)$.  The factor
$[\Lambda({\bf r})/\epsilon]^2$ is a dimensionless measure of the
tunneling coupling between the bare states in the two wells. It can be
seen that $\tau$ has a minimum with respect to $\Lambda$ when
$\Lambda= \epsilon$, which is equivalent to $\Delta=0$. This condition
defines the minimal relaxation time $\tau_0$ referred to in
Eq.~(\ref{GammaRel}).  The minimal $\tau$ corresponds to the symmetrical 
configuration when the bare energy levels at both sites are equal.  To
allow for time dependence of $\tau$ one should substitute $\epsilon =
\epsilon(t)$ from Eq.~(\ref{esplitting}).

The expression (\ref{rt}) contains two specific energy scales,
$E_a \equiv 2 \hbar s/a$, and $E_r \equiv \hbar s/r$. The first scale is
the energy of a phonon having a wavelength of the order of the
single-site localization length, $a$. Phonons with larger energies
produce rapidly oscillating fields which average out at the
distance occupied by a localized electron. Consequently, $\tau^{-1}$
strongly decays at $\epsilon \gtrsim E_a$. The second scale
corresponds to phonons with a wavelength of the order of the
distance $r$ between the components of the pair. If the deformation
potentials of both components of the pair are the same, or if the main
mechanism of the electron-phonon interaction is piezoelectric, then at $2
\epsilon \ll E_r$ both levels move synchronously, and no
interaction occurs. The net interaction is hence proportional to some
power of the ratio $2 \epsilon/E_r$, see Ref.~\onlinecite{GGPRev} for a review. 

In most realistic cases one can assume $\epsilon \ll E_a$.  However,
 for $T\approx 1$ K the ratio $x=2\epsilon/E_r$ can be either less or
 greater than 1, giving different types of behavior for $\Phi_n(x)$.
 We will concentrate on the case of $x \gg 1$ since this limiting case
 seems to be more easily accessible for experiments. In this regime,
 $\Phi_n(x)$ can be considered as constant. The validity of this
 approximation will be considered in the Discussion. The quantities
 $\tau_n (T)$ are listed in Ref.~\onlinecite{GGPRev}.

The most important feature of the relaxation for our problem is the
\emph{energy dependence} of the relaxation rate, namely, the power $n$.  
Under the above-mentioned conditions, $n=3$ in the case of deformational
interaction and $n=1$ for piezoelectric interactions.\cite{GGPRev} As will
be clear later, only the energies $\epsilon \ll k_B T$ are important
for the anomalous nonlinear behavior, so one can approximate $\coth
(\epsilon/k_B T) \approx k_BT/\epsilon$. In this way we arrive at the
following energy dependences of the relaxation rate:
\begin{equation} %\label{}
\frac{1}{\tau} \propto \left\{ \begin{array} {ll}
\Lambda^2 \epsilon^0 & \text{for the deformational interaction}\\
\Lambda^2 \epsilon^{-2} & \text{for the piezoelectric interaction}
\end{array} \right.
\end{equation}
Note that in the first case the relaxation time 
is independent on $\epsilon$ and thereby on time. This
 is the same as is the case in dielectric glasses. Apart from a 
constant and the dependency on the magnetic field, we can thus expect 
the same type of behavior from these two very different systems.
In the second case the $\epsilon$-dependence is the same as in
metallic glasses, which has also been analyzed by the authors\cite{KG} and has
proven to be the source of a pronounced
anomalous effect. In particular, in metallic glasses the lowest
nonlinear contribution is proportional to the 
intensity to $3/2$ rather than to the intensity squared as for
dielectric glasses. 

In the absence of the magnetic field the energy overlap integral
$\Lambda$ is related to the distance $r$ between the sites of a pair
simply as $\Lambda=\Lambda_0 \exp^{-r/a}$, where $\Lambda_0 =
(1-5)\times me^4/ \hbar^2 \kappa^2$ is of the order of the effective
Bohr energy. Here $m$ is the electron effective mass while $\kappa$ is
the dielectric constant.  A magnetic field will squeeze the electron
wave function, and this effect will be strongest for the direction
perpendicular to the field. This introduces an angular dependency to
the localization length and thereby also to $\Lambda$.  Following
Ref.~\onlinecite{ESMag}, we will analyze the limiting cases of weak
(w) and strong (s) magnetic field where the influence of magnetic
field is weak or strong, respectively. The asymptotic expressions for
the $\zeta ({\bf r})\equiv- \ln [\Lambda ({\bf r})/\Lambda_0]$ are the
following\cite{ESMag}:
\begin{eqnarray}
\zeta_w&=&\frac{r}{a}+\frac{r^3a \sin^2 \theta }{24
{\lambda}^4}\, ,\\ 
\zeta_s&=&\frac{r^2 \sin^2 \theta}{4 {\lambda}^2}+\frac{|r \cos
\theta|}{a_H}\, .
\end{eqnarray} 
Here $\theta$ is  the angle between $\bf r$ and the direction of
the magnetic field $\bf H$, ${\lambda}=\sqrt{ \hbar c/e H}$ is the
magnetic length, while
$a_H=\hbar / \sqrt{2 m E_H}$ is the characteristic localization
length in the longitudinal direction, where $E_H$ is the ionization energy of
the ground state of the localized electron in the magnetic field.

\subsection{Pair distribution function}

The total absorption is given by a sum of the contributions of the
individual pairs. Hence we have to sum over the $\bf r$, as well as
over the individual energies of the electron levels.  
The latter
summation must take into account the correlation between the level
occupation numbers due to Coulomb interaction. As shown in
Ref.~\onlinecite{ESCoul}, the summation over the energies can be split
into integration over the pair center-of-gravity and over the bare
inter-level spacing $\Delta$. The first integration gives $2 \Delta+ e^2/\kappa
r$ since only the pairs with the center-of-gravity energy between the
chemical potential $\mu+ \Delta$ and $\mu -\Delta - e^2/\kappa r$ have one electron
per pair.  As a result, the pair distribution function
can be expressed through the single-electron density of states $g$
as\cite{ESCoul},    
\begin{equation}\label{rtrDist}
N(\Delta,{\bf r})=\frac{g^2 V}{4\pi}\left(2 \Delta+\frac{e^2}{ \kappa
r} \right)
\end{equation}
where $V$ is the volume contributing to the absorption. This
expression is valid if for a typical hopping distance $e^2/\kappa r
\gg \Delta_C$ where $\Delta_C$ in the width of the \emph{Coulomb gap}
in the single-electron density of states.\cite{ESCoul} Inside the
Coulomb gap, we would rather have to use the distribution
\begin{equation} \label {dfc}
N(\Delta,{\bf r})=\frac{3}{40 \pi^3} \left(\frac{
\kappa}{e^{2}}\right)^6\left(2 \Delta +\frac{e^2}{ \kappa r}\right)^5\, .
\end{equation}
Calculations for both cases are similar.  Note that the distribution
function is isotropic. Under the conditions of interest to us the
typical hopping distance $r$ is small enough to let us neglect $2
\Delta$ in comparison with $e^2/\kappa r$ in Eqs.~(\ref{rtrDist})
and~(\ref{dfc}). Thus the distribution becomes $\Delta$-independent,
and we denote it as $N({\bf r})$. Using the notation presented above
we can now write the coupling constants $\alpha_1,alpha_2\approx \alpha$ from
Eqs. \ref{GammaRes}, and~\ref{GammaRel} as\cite{GalPriev}
\begin{equation}\label{alpha}
\alpha=\frac{4 \pi^3}{3}{\cal K}^2 \frac{e^4 g^2 a r_{\omega}^3}{\kappa^{2}},\quad r_{\omega}=a \ln{\frac{\Lambda_0}{\hbar \omega}},
\end{equation}
where ${\cal K}$ is the coupling constant of the piezoelectric
interaction. The power of $r_{\omega}$ may vary for different type of
interactions, depending on whether the interaction includes the dipole
moment of the pair.

\section{Calculation of absorption} \label{ca}
As a result of the previous considerations, the absorbed power can be
expressed as  
%\begin{eqnarray} 
\wide \begin{equation}
P =
%&=&% 
\int d\bn \int_0^\infty \!d \Delta  \int_0^\infty \! r^2\, dr \,  N(r) 
 \left\{ 
%\, p(\Delta,r,\bn) \, ;\label{semiabs}\\
%p(\Delta,r,\bn) &=&
\int_0^{\Theta} \!\int_0^{\Theta}\! \frac{dt\,
dt'}{\Theta \, k_B T}\,  \frac{
\,{\dot \epsilon} (\br,t)\, {\dot
\epsilon}(\br,t-t')
}{ \cosh^{2} [\epsilon (\br,t-t')/k_B T]}
%\nonumber \\ &&\qquad \times 
\, \frac{\exp\left(- \int_0^{t'} dt_1/\tau
(t-t_1) \right)}{1 - \exp [ -\int_0^{\Theta}\!
dt_1/\tau (t_1)]}\right\} \, . \label{semiabs}
%\end{eqnarray}
\end{equation} \narrow
Here the expression in the braces is just the power absorbed by an
individual pair, $p(\Delta, \br)$. The  
energy splitting $2 \epsilon$ depends on $\br$ through the
interaction potential $d_0(\br)$ given by Eq.~(\ref{ip2}) and through the
tunneling splitting 
$\Lambda= \Lambda_0\, e^{-\zeta(\br)}$. Technically it is convenient
to transform the integral from the set of variables $r,\bn$,
to the variables $\Lambda,\theta, \phi$. Such a
transform introduces the factor
$$
\left|\frac{\partial \Lambda (r,\bn) }{\partial
r}\right|^{-1}=\frac{a}{\Lambda} f(\Lambda,\bn)\, .$$
Here the dimensionless function $f(\Lambda,\bn)$ is given by the
equation
\begin{equation} \label{Jac1}
f(\Lambda,\bn) \equiv - \frac{1}{a}
\frac{\partial r_\Lambda(\bn)}{\partial \ln \Lambda} =\frac{1}{a}
\frac{\partial r_\Lambda(\bn)}{\partial {\cal L}}\, .
\end{equation} 
where ${\cal L} \equiv \ln (\Lambda_0/\Lambda)$, while
$r_\Lambda(\bn)$ is the solution of the equation  
\begin{equation} \label{rl}
\zeta (r_\lambda, \bn)= {\cal L} \, .
\end{equation}
In the simplest case of zero magnetic field $f=1$, and we have a
distribution very similar to those of glasses. The variable transform
strongly simplifies the calculations since the relaxation time is a
function of the parameters $\Delta$ and $\Lambda$ and the quantity
$r_\Lambda$ is a weak (logarithmic) function of
$\Lambda$. Consequently it can be extracted out of the integral over
$\Lambda$, while replacing $\Lambda$ in the expression for $r_\Lambda$ 
by its
characteristic value. As a result, the integrations over $\Delta$ and
$\Lambda$ will remain the same as previously calculated for glasses,
and only the angular integration and the dependence on the 
characteristic value of $\Lambda$
are different.

The following calculation procedure is similar to that of
Ref.~\onlinecite{GalPriev}. The expression (\ref{semiabs}) will be
expanded in powers of the effective electric field, and the lowest
correction will be compared with the linear result. Expanding the
individual contributions in powers of $d_0$ as 
\begin{equation} %\label{}
p(\Delta, \br)=\sum_{k=2}^4 p^{(k)} d_0^k (\br)
\end{equation}  
we notice the the coefficients $p^{(k)}$ depend only on the quantities
$\Delta$ and $\Lambda$. Transforming the variables from $r,\bn$ to
$\Lambda, \bn$ we can use the fact that $r$ is a weak function of
$\Lambda$ and extract of the quantities proportional to the powers of
$r$ from the integral over $\Lambda$ replacing
$$r \rightarrow r_c (\bn)=r_\Lambda
(\bn)\left|_{\Lambda=\Lambda_c}\right.$$ where $\Lambda_c$ is the
characteristic value determined by the integrand over $\Lambda$. In a
similar way, we replace $f(\Lambda, \bn) \rightarrow
f_c(\bn)=f(\Lambda_c, \bn)$.  Finally we arrive at the expression
$P=\sum_{k=2}^4 P^{(k)}$ with $P^{(k)}=(a e{\cal E}_0)^kI_kJ_k$ where
\begin{eqnarray} \label{finabs}
I_k&=&\int_0^\infty \! d \Delta \int_0^\infty \! d \Lambda\, \Lambda^{-1}p^{(k)}(\Delta,
\Lambda)\, , \label {ik}\\
J_k&=&\int d\bn\,(\bbox{\nu}\cdot \bn)^k f_c\, N(r_c)\, (r_c/a)^k \label{jk} 
\end{eqnarray} 
The quantities $I_k$ are the same that enter the expressions for
nonlinear absorption in glasses, and we quote them from Ref.~\onlinecite{KG}.

Parallel with $I_k$ we can also estimate $\Lambda_c$, and thereby
${\cal L}$. It can be shown that this value is only weakly dependent on
the mechanism of absorption and on $k$. For all the mechanisms analyzed 
in this paper we have
\begin{equation}
{\cal L} 
=\ln \frac{c_{\cal L}\Lambda_0}{k_BT \sqrt{\omega \tau_n(T)}}.
\label{call}
\end{equation}
The value for $c_{\cal L}$ is of the order one, and varies only with a
factor less than two for the situations discussed in this paper. We will
therefore treat ${\cal L}$ as independent of $k$ for this discussion.

In this paper we will present the results for the case of low
frequencies, $\omega \tau_n \ll 1$, for which the anomalous nonlinear
behavior is most pronounced. In this case the linear in the intensity
contribution, $I_0$, is independent of $\tau$, and moreover, it is
the same for any dependence $\tau (\epsilon)$,
 \begin{equation} \label{lc}
I_0 =(\pi^2/16)\, \omega \approx 0.62 \, \omega \, . 
\end{equation}

The case of deformational interaction, for which the parameter $n$ in
Eq.~(\ref{rt}) is equal to 3, corresponds to the situation in
dielectric glasses. In this case\cite{KG}
\begin{equation}\label{n3}
I_3=0\, , \quad I_4 \approx\\
0.054\,\frac{\omega}{\sqrt{\omega \tau_3(T)}(k_BT)^2} \, .
\end{equation}

The piezoelectric interaction ($n=1$) is
similar to the case of metallic glasses. It can be shown that the
energy dependence of the relaxation time leads to a divergence in the
integration over $\Lambda$ and $\Delta$. To get a proper estimate one
should cut off the integration at $\epsilon \lesssim d_0$. As a
result, the leading nonlinear contribution appears proportional to
$|d_0|^3$ and equal to\cite{KG} 
\begin{equation}
I_3 \approx 0.1 \,\omega/k_BT.
\end{equation}
We again expect ${\cal L}$ to be given by Eq. \ref{call}.  To
calculate the angular integrals $J_k$ we solve Eq.~(\ref{rl}) in a
recursive way.  For the case of weak magnetic fields, we obtain
\begin{eqnarray}
r_c &=&a {\cal L} \left(1- {\cal L}^2 a^4 
\sin^2 \theta/24\lambda^4\right)\, ; \label{rlw} \\
f_c&=&\
1+a^4 {\cal L}^2 \cos^2 \theta/8\lambda^4\, . \label{fw}
\end{eqnarray}
For the strong field limit, solving a quadratic equation, we obtain
\begin{eqnarray}
r_c  &=& \frac{2 \lambda^2 \cos \theta}{ a_H\sin^2
\theta} \left( 
\sqrt{1+ \tan^2\theta {\cal L}(a_H/\lambda)^2} -1 \right) \, ;
\label{rls}\\
f_c &=&(a_H/a){\cal L} \cos \theta  \left(1+
{\cal L}(a_H/\lambda)^2\tan^2\theta \right)^{-1/2} .\label{fs}
\end{eqnarray}
Here $\theta$ is the angle between $\bn$ and the direction of magnetic
field $\bf H$. Hence, $d\bn = d(\cos \theta) \, d\phi$ where $\phi$ is
the azimuthal angle between the projections of $\bn$ and $\bbox{\cal
E}_0$ on the plane, perpendicular to $\bf H$. After substitution of
Eqs.~(\ref{rlw})--(\ref{fs}) into Eq.~(\ref{jk}) the angular integrals
are calculated directly.

\section{Results}

%For results for the linear terms we refer to previous
%publications\cite{GGPRev,GalPriev}, here we will only give the
%non-linear results.  

To set the scale of nonlinear corrections let us start with the
expression for the linear absorption in the absence of magnetic
field, $P_0 \equiv P_0(0)$. 
Both
linear and nonlinear contributions are dependent on the relaxation
mechanism of the relevant pairs.
%We will first consider deformational interactions,
%interacting through a dipole moment. This gives us a linear absorption of
\subsection{Deformational interaction between localized pairs and
phonons}
We will first consider deformational interaction between the localized
pairs and thermal phonons.
For the deformational mechanism, 
\begin{equation}\label{P20}
P_0=(\pi^2/48) \,(V a^4 g^2 e^4 \omega {\cal
L}^3{\cal E}_0^2 / \kappa)\, . 
\end{equation}
In the absence of magnetic field we obtain
\begin{equation}
P_4(0)=\frac{P(0)-P_0(0)}{P_0(0)} =  0.26 \frac{F^2 {\cal
L}_c^2}{ \sqrt{\omega\tau_0(T)}} \label{P40}
\end{equation} 
were we have introduced the dimensionless
``field amplitude'' 
\begin{equation} 
F \equiv e {\cal E}_0 a/k_B T \, .
\label{0aa}
\end{equation}
In  a weak magnetic field, the quadratic in magnetic field corrections
arise both to the linear absorption and to the lowest nonlinear
contribution. They can be expressed in a unified way as, cf. with
Ref.~\onlinecite{GalPriev},
\begin{equation}\label{Pw}
P_{0/4}^{(w)} (H)=P_{0/4}(0) \left[1-c_w (a /\lambda)^4{\cal L}^2\right]\, , 
\end{equation}
so the magnetic field produces corrections which are $\propto H^2$. 
The numerical factor $c_w$ depends on the direction of the electric field
$\bbox{\cal E}_0$ with respect to the magnetic field $\bf H$. Its
values are also different for the linear and nonlinear,
contributions, $c_w^{(0)}$ and $c_w^{(4)}$, respectively. 
The values of $c_w$ are shown in Table I. 
\begin{center}
\begin{minipage}{3in}
\begin{center}
\begin{tabular}{|l|l|l|l|} \hline
Direction&$c_w^{(0)}$ &$c_w^{(4)}$  &$c_w^{(3)}$ \\ \hline
${\bf H} \perp \bbox{{\cal E}}_0$ & 0.2   & $\approx 0.29$ & $\approx 0.25$ \\
 ${\bf H} \parallel \bbox{{\cal E}}_0$ &0.1 &$\approx 0.095$& $\approx 0.097$
\\ \hline
\end{tabular}

\medskip
Table I. Numerical coefficients entering the nonlinear contributions
to the absorption.
\end{center}
\end{minipage}
\end{center}

 The decrease of attenuation in the
magnetic field has the following physical reason. The presence of a
magnetic field squeezes the electron wave functions, and the overlap
integrals between the components of the pair decrease. Furthermore,
the wave functions are squeezed mostly in the direction perpendicular
to $\bf H$. On the other hand, the coupling between the wave and the
pair is maximal if the pair dipole moment is parallel to $\bbox{\cal
E}_0$.  Thus the reduction of absorption is more pronounced for ${\bf
H}\perp {\bf n}$. 
For the nonlinear contribution the difference should be
even stronger, as it includes higher orders of the dipole moment. 

In the limit of strong magnetic fields the results are even more
interesting, as the functional dependency on the magnetic field also
varies with the different absorption types. Yet we still see the same
relative considerations as for the weak field limit. For a magnetic
field parallel to the radiation polarization vector we get
\begin{eqnarray}\label{P4sa}
\frac{P_0^{(s)}(H)}{P_0}&=&3\frac{\lambda^2 a_H^2}{{\cal L} a^4}
\propto H^{-4/3}\, ,\\ 
\frac{P_{4}^{(s)}(H)}{P_{4}(0)}&=&\frac{5}{2}\frac{\lambda^2 a_H^4}{{\cal L}a^6} \propto
H^{-5/3}\, . 
\end{eqnarray}
For a perpendicular field the results show both a stronger dependency
on $H$ and on the order of the expansion in intensity.
\begin{eqnarray}\label{P4se}
\frac{P_0^{(s)}(H)}{P_0(0)}&=&6\frac{\lambda^4} {{\cal L}^2 a^4}
\ln{\frac{{\cal L}a_H^2}{\lambda^2}} \propto H^{-2}\ln{H}\, ,\\
\frac{P_{4}^{(s)}(H)}{P_{4}(0)}&=&30\frac{\lambda^6}{{\cal L}^3 a^6}\ln{\frac{{\cal L}a_H^2}{\lambda^2}}
\propto H^{-3}\ln{H}
\end{eqnarray}

\subsection{Piezoelectric interaction between localized pairs and
phonons} 

We will now turn our attention to the piezoelectric interaction. This
also interacts via a dipole moment, so the linear results are
basically the same, apart from a coupling constant.
However, there is a striking difference for the nonlinear
contribution -- similarly to the case of metallic glasses,\cite{KG} 
the integration over $\Delta$ and
$\Lambda$ in Eq.~(\ref{ik}) results in a term proportional to
$|d_0|^3$ rather than $d^4_0$. This is reflected in a change of the
absorption dependencies both on the wave intensity and on the magnetic field. 
For this case, we restrict ourselves by order-of-magnitude estimates
for numerical factors. Calculation of exact numbers would require a
great amount of numerical work which would be inadequate to the
accuracy of the initial model for the electron density of states. 
The results read as,
\begin{equation}\label{P30}
P_3(0)=[P-P(0)]/P_0=  c_m |F|\, ,
\end{equation}
where the %order-of-magnitude 
estimate for $c_m$ is 0.1. 
Note that in this case the expansion of the absorption  in powers of
intensity appears \emph{non-analytical} which implies relatively
strong nonlinearity.

Since now the expansion in $d$ of the nonlinear contribution starts with
$|d_0|^3$ rather than from $d^4_0$ one can expect weaker magnetic field
effects. This is indeed the case. The values of the numerical
coefficients $c_w^{(3)}$ are shown in the Table I. 

For strong fields we get for parallel and perpendicular fields respectively
\begin{eqnarray}\label{P3s}
\left(\frac{P_{3}^{(s)}(H)}{P_{3}(0)}\right)_\parallel&=& \frac{8}{3}\frac{\lambda^2 a_H^3}{{\cal L} a^5}\propto H^{-3/2}\\
\left(\frac{P_{3}^{(s)}(H)}{P_{3}(0)}\right)_\perp&=&
\frac{64}{3}\frac{\lambda^5 }{{\cal L}^{5/2} a^5}\ln{\frac{{\cal L}a_H^2}{\lambda^2}}\propto H^{-5/2}\ln{H}.
\end{eqnarray}
Let us recall that the above expressions are valid for the low
frequency limit only, where $\omega \tau_n(T)<<1$. Similar
calculations are possible for the high frequency limit, as well as for
different pair distribution functions. In particular, for the case of
pronounced Coulomb gap the essential differences occur only in powers
of  $r$ and thereby of $\cal L$. Thus
the influence of the magnetic field will be different. The relation
between linear and nonlinear results remain similar, apart from
numerical factors. 

\section{Discussion}

Let us discuss the relevance of the obtained results for realistic
materials and situations. In this connection, 
 several parameters are to be considered.

Regarding the material properties, we consider only \emph{weakly
doped} or amorphous semiconductors in the regime of
\emph{nearest-neighbor hopping conductance}. Hence, we calculate
absorption by close pairs independent of each other. To keep the model
adequate we have to require that the typical inter-center distance
within the pair, $r_c$, should be much smaller that the typical
distance between defect centers, ${\bar r} =(4\pi n_d/3)^{-1/3}$. Here
$n_d$ is the defect concentration. The
\emph{hopping distance} $r_c$ is discussed in Sec.~\ref{ca}.  

Another requirement is that  the impurities are not too shallow, so that the
electrons cannot be excited from the localized states to the
conduction band by the AC perturbing potential. There are experimental
examples of this, where such a excitation serves as a source of
nonlinear behavior.\cite{GGDL}

According to the present calculation, the most interesting effects
occur at ``low'' frequencies when $\omega \tau_n (T) \ll
1$. 
This requirement also ensures that the relaxation absorption 
dominates the resonant one.
Certainly, the minimal relaxation time $\tau_n(T)$ is a material
property. Usually the above inequality is met at low temperatures for
frequencies in the range $100-1000$ MHz. 

The TLS model in glasses is restricted to very low temperatures where
higher energy levels are not excited. The situation is a bit different
in semiconductor materials where the inter-level splittings are of the
 order of the Bohr energy. Consequently, the nearest-neighbor hopping
conductance can be effective in the temperature range up to a few K.

The main objectives of this paper is to show that nonlinear
effects are anomalously large. Indeed, in the case of deformational
absorption an additional parameter $(\omega \tau_3)^{-1/2} \gg 1$ is
present in the nonlinear expansion (\ref{P40}), while the the case of
piezoelectric interaction the nonlinear expansion starts from the
\emph{fist power} of dimensionless amplitude $|F|$, Eq,~(\ref{P30}).
Furthermore, the nonlinear contributions have pronounced magnetic field
dependences different from the linear ones. We hope that those
features will allow experimentalists to detect the nonlinear behavior 
and to discriminate between different relaxation mechanisms for
localized states. 

In course of the present calculations we have assumed the inequality
$\epsilon/E_r\gg1$ to be met, see Eq.(\ref{rt}).  It is important that
for $\epsilon$ one has to substitute the value which gives the
dominating contribution in the final integration over the pair
distribution function. This value is actually different for the linear
and the nonlinear contributions, and it is also dependent on the
relaxation mechanism.  It turns out that for the linear absorption
this typical $\epsilon \sim k_B T$, while for the nonlinear
contributions it is reduced by a factor $\sqrt{\omega \tau_3}$ for the
deformational interaction, by $|F|$ for the piezoelectric one, both
calculated under the condition $\epsilon \gg E_r$.  A similar estimate
is necessary to choose a proper value for $r$ in the expression for
$E_r$. This value depends on the quantities $\sqrt{\omega \tau_n}$,
$\Lambda_0$, as well as on the magnetic field.  Thus the experimental
variables intensity, frequency and magnetic field, in addition to the
system parameters $\tau_0$ and $\Lambda_0$, influence the behavior of
the nonlinear absorption. This rich parameter space allows for a large
range of experiments.

\section{Conclusions}

In this paper, we have analyzed nonlinear contributions to the
acoustic and electromagnetic absorption by localized electron states
in semiconductors in the regime of hopping conductance. 
The most important conclusions are the following.
\begin{itemize}
\item The total behavior of absorption is determined by the product
$\omega \tau$ where $\tau (T)$ is the minimal relaxation time for
a pair with energy splitting of the order $k_B T$. 
\item The anomalous nonlinear behavior occurs at $\omega \tau \ll 1$.
In the case of deformational relaxation mechanism for the localized
electrons  a large  additional factor $(\omega
\tau)^{-1/2}$ appears in front of  the item $\propto F^2$ in the expansion of
nonlinear absorption. In the case of piezoelectric  relaxation
mechanism  the expansion starts with the item $\propto |F|$ rather
than $\propto F^2$. Here $F$ is the dimensional AC field amplitude.
\item The anomalous nonlinear absorption is strongly influenced by an
external magnetic field, the influence being dependent both on the
electron pair distribution function, on the 
dominating relaxation mechanism for the localized electrons, and  on the
direction of the field polarization vector with respect to the
magnetic field. The influence of magnetic field on the linear
absorption and nonlinear corrections is substantially different.
\end{itemize}
As a result, the physical picture of weakly nonlinear absorption
appears rich and informative. Our estimates show that the effects
under consideration are accessible for the modern experiment, and many
important characteristics -- the dominating relaxation mechanism,
the importance of the Coulomb gap, typical hopping distances, etc. --
can be extracted by comparison to the present theory provided the
experiment will be done. 

It should be emphasized that there is a close similarity between the
present and the results of our previous calculations for glassy
materials.\cite{KG} However, the localized states in disordered
semiconductors, being charged, can be influenced by magnetic field
which makes them easier to investigate. We therefore also hope that
the studies of semiconductor systems will also provide a new information
regarding nonlinear response of TLSs in glasses.

\widetext

\begin{thebibliography}{99}
\bibitem{PollGeb} M. Pollack and T. H. Geballe, Physical Review {\bf 122 6}, 
1742 (1961).
\bibitem{GGPRev}Yu. M. Galperin, V. L. Gurevich and  D. A. Parshin,
Pis`ma  Zh. Eksp.  Teor. Fiz. {\bf 36}, 386 (1982)
    [JETP Lett. {\bf 36}, 466 (1982)];
Zh. Eksp. Teor. Fiz. {\bf 86}, 1900 (1984)  [Sov. Phys. JETP {\bf 59},
1104 (1984)].
\bibitem{Jackle} J. Jackle, Z. Phys. {\bf 257}, 212 (1972).
\bibitem{ESAbs}  A. L. Efros and B. I. Shklovskii, in {\em
Electron-Electron Interactions in Disordered Systems}, edited by
A.L. Efros and M. Pollak (Elsevier, Amsterdam 1985) p. 201. 
\bibitem{Long} A. R. Long, Adv. Phys. {\bf 31} 553 (1982).
\bibitem{GGDL}Yu. M. Galperin, E. M. Gershenzon, I. L. Drichko and
L. B. Litvak-Gorskaya, Sov. Phys. Semicond. {\bf 24} 1 (1990).
\bibitem{AHV} P. W. Anderson, B. I. Halperin and C. M. Varma,
Philos. Mag. {\bf 25}, 1 (1972).
\bibitem{Phil} W. A. Phillips, J. Low. Temp. Phys. {\bf 7}, 351
(1972). 
\bibitem{KG} M. Kirkengen and Y. M. Galperin, Phys. Rev. B {\bf 56}
13615 (1997).
\bibitem{Galdiel}Yu. M. Galperin.  Zh. Eksp. Teor. Fiz. {\bf 85}, 1386
[Sov. Phys. JETP, {\bf 58}, 804 (1983)].
\bibitem{HA} S. Hunklinger and W. Arnold, in {\em Physical Acoustics},
edited by W. P. Mason and R. N. Thornton (Academic New York 1976), 
Vol. XII p. 155.
\bibitem{Black} J. L. Black, in {\em Glassy Metals I}, edited by
H. J. G\"unterodt and H. Beck (Springer, Berlin 1981), p. 245. 
\bibitem{end1} 
In the literature one finds the energy splitting variously defined as
$\epsilon$ or $2\epsilon$. Thus some equations may appear differently
here than in the sources from wich they are quoted.
\bibitem{end2} The case of deformation interaction neads a bit
different treatment. However, the results are very similar.
\bibitem{ESMag} B. I. Shklovskii and A. L. Efros, {\em Electronic
properties of Doped Semiconductors} (Springer, Berlin 1984). 
\bibitem{ESCoul} A. L. Efros and B. I. Shklovskii, in {\em
Electron-Electron Interactions in Disordered Systems}, edited by
A.L. Efros and M. Pollak (Elsevier, Amsterdam 1985) p. 409. 

\bibitem{GalPriev}Yu. M. Galperin and
E. Ya. Priev. Fiz. Tverd. Tela(Leningrad), {\bf 28}, 692 (1986). 
%\bibitem{MG} P. Doussineau, J. Phys. (France) Lett. {\bf 42}, L83
%(1981); H. Araki {\em et al.}, Phys. Rev. B {\bf 21}, 4470 (1980);
%G. Park {\em et al.}, Phys. Rev. B {\bf 24}, 7389 (1981); A. Hikata
%{\em et al.}, J. Low. Temp. Phys. {\bf 49}, 339 (1982). 




\end{thebibliography}
\end{document}